\documentclass[pdflatex,sn-basic,Numbered]{sn-jnl}

\usepackage{graphicx}%
\usepackage{multirow}%
\usepackage{amsmath,amssymb,amsfonts}%
\usepackage{amsthm}%
\usepackage{mathrsfs}%
\usepackage[title]{appendix}%
\usepackage{xcolor}%
\usepackage{textcomp}%
\usepackage{manyfoot}%
\usepackage{booktabs}%
\usepackage{algorithm}%
\usepackage{algorithmicx}%
\usepackage{algpseudocode}%
\usepackage{listings}%

\usepackage{tikz}
\usepackage{adjustbox}
\usetikzlibrary{positioning, arrows.meta}
\usepackage{booktabs}
\usepackage{siunitx} 
\usepackage{xcolor}
\usepackage{arydshln}
\usepackage{tabularx}
\let\orcidlogo\relax
\usepackage{orcidlink}
\usepackage{siunitx}

\theoremstyle{thmstyleone}%
\theoremstyle{thmstyletwo}%

\theoremstyle{thmstylethree}%

\begin{document}

\title[Human-Centered Semantic Validation of Network Traffic Classification]{Beyond Measurement Metrics: A Human-Centered Framework for Semantic Validation of Network Traffic Classification}


\author*[1]{\fnm{Igor} \sur{Cherepanov}
\orcidlink{0000-0001-9109-090X}
}
\email{igor.cherepanov@igd.fraunhofer.de}

\author[1]{\fnm{David} \sur{Sessler}
\orcidlink{0009-0006-0258-7927}
}

\author[1]{\fnm{Alex} \sur{Ulmer}
\orcidlink{0009-0008-3778-8184}}
\author[1]{\fnm{Thorsten} \sur{May}
\orcidlink{0000-0001-8027-2687}}
\author[1,2]{\fnm{Jörn} \sur{Kohlhammer}
\orcidlink{0000-0003-1706-8979}}

\affil[1]{\orgname{Fraunhofer IGD}, \orgaddress{\city{Darmstadt}, \postcode{64283}, \country{Germany}}}

\affil[2]{\orgname{Technische Universität Darmstadt}, \orgaddress{\city{Darmstadt}, \postcode{64289}, \country{Germany}}}


\abstract{Machine learning (ML) has become the dominant approach for network traffic classification, achieving very high predictive performance. 
However, a model is only valuable if it learns semantically meaningful and trustworthy patterns rather than exploiting spurious correlations.
Conventional evaluation practices predominantly assess predictive performance. 
Consequently, whether the model relies on semantically meaningful patterns remains unknown.
To address these challenges, we adapt the knowledge generation framework for network traffic classification. 
The adapted framework combines data, ML models, explainability, visualization, and expert reasoning to support the iterative exploration, verification, and refinement of model behavior and data preprocessing.
The framework is grounded in findings from the literature, benchmark dataset analyses, practical experience with XAI-based traffic classification, and expert feedback, providing practical guidance for semantic model validation.
By complementing predictive performance with semantic validation and human expertise, the proposed framework supports the development of network traffic classification models that are not only accurate but also robust and trustworthy.}


\maketitle
    
\section{Introduction}\label{intro}

Network traffic classification is essential for identifying the applications and services generating network traffic, enabling effective network management, security monitoring, anomaly detection, and quality-of-service provisioning.
Over time, traffic classification techniques have evolved from simple port-based approaches, which identify applications based on well-known transport-layer port numbers, and statistical methods to increasingly sophisticated ML and deep learning (DL) models~\cite{4738466, Salman2020}. 
Recent advances in these methods have significantly improved predictive performance, leading research efforts to focus primarily on the development of more accurate classification models.

While existing evaluation practices provide insights into how well a model performs, they provide limited support for understanding why a model performs well. 
Consequently, models can achieve high predictive accuracy, while they may rely on dataset-specific shortcut features, spurious correlations, protocol-specific artifacts, dataset biases, or other unintended patterns unrelated to the underlying classification task~\cite{Geirhos2020-aw}.
Such dependencies often remain hidden during development and may lead to substantial performance degradation when models are deployed in evolving network environments, where traffic characteristics and data distributions continuously change~\cite{8543584,9829791}.

Current traffic classification pipelines predominantly assess predictive performance. 
They provide little insight into what the model has actually learned. 
Semantic validation addresses this limitation by determining whether the learned decision strategy is based on meaningful traffic characteristics.
At the same time, research fields such as explainable artificial intelligence (XAI), human-centered artificial intelligence (HCAI), and visual analytics (VA) increasingly emphasize the importance of understanding, validating, and contextualizing model behavior through human expertise~\cite{10.1145/3290605.3300233, 10.1145/3561048, analytics4010007, 11033407}. 
Despite these advances, systematic approaches for integrating explanation-based validation and expert reasoning into network traffic classification remain largely unexplored.

Motivated by this gap, this work builds upon the knowledge gained from our previous research on explainable traffic classification, including the design, implementation, and evaluation of human-centered VA system for explanation analysis and distribution shift assessment~\cite{grivapp26}. 
That work highlighted that robust traffic classification requires more than accurate prediction.
Models must remain reliable under evolving network conditions, while enabling experts to understand, validate, and improve their behavior. 
The practical experience obtained from that work revealed recurring challenges that extend beyond individual explanation techniques and highlighted the need for a systematic framework that integrates explainability, expert reasoning, semantic model validation, and robustness assessment throughout the ML lifecycle. 
Ultimately, the goal is to actively involve domain experts in the continuous evaluation and refinement of ML models, supporting traffic classifiers that remain robust, reliable, and trustworthy.
The resulting insights may reveal deficiencies in the learned decision strategy, thereby informing revisions to earlier stages of the pipeline.
Consequently, explainability is viewed not merely as a means of understanding model behavior, but as a mechanism for continuously improving data, feature representations, models, and evaluation strategies.

To support semantic model validation in network traffic classification, we adopt the knowledge generation model for VA proposed by Sacha et al.~\cite{6875967}, as it already provides the core stages of iterative, human-centered knowledge generation. We adapt the framework by incorporating explainability, stakeholder-specific objectives, and explanation-based evaluation into the workflow. The resulting framework extends conventional performance-centric evaluation with semantic model validation, supporting the development of traffic classification models that are not only accurate but also transparent, robust, and correct for the right reasons. 
Overall, our contributions are:

\begin{itemize}
\item We identify recurring shortcomings in current network traffic classification workflows, including dataset artifacts, performance-centric evaluation, and limited integration of explainability and expert reasoning, and translate these findings into a practical framework for trustworthy model development.

\item We apply the knowledge generation model for VA into a human-centered framework for explainable traffic classification that connects user, tasks, goals, explanation requirements, and evaluation strategies to support trustworthy model development and assessment.

\item The resulting framework establishes a structured design space for human-centered XAI in traffic classification by organizing explanation generation and evaluation according to user roles, analytical goals, explanation levels, explanation requirements, and semantic validation criteria.

\item Based on the proposed framework, we derive design recommendations for trustworthy XAI evaluation by incorporating domain experts and semantic validation.

\end{itemize}

\section{Related Work}

Network traffic classification has evolved considerably over the past decades. 
Early approaches relied on well-known port numbers and deep packet inspection (DPI), where applications were identified through predefined ports or protocol signatures. 
However, port-based classification became unreliable because applications increasingly employed dynamic and shared ports, while the growing adoption of encryption substantially reduced the effectiveness of DPI by concealing application-layer payloads~\cite{4738466}. 
These limitations motivated a shift toward ML-based traffic classification. 
Initial research employed classical supervised learning algorithms, such as decision trees, naïve Bayes classifiers, support vector machines, k-nearest neighbors, random forests, and ensemble methods, which significantly improved classification performance compared with port-based classification and DPI.

As larger benchmark datasets became available and computational resources advanced, deep learning (DL) emerged as the dominant paradigm, shifting the focus from manually engineered features to representation learning~\cite{AZAB2024676, salman2020review, 8669045}. 
Neural networks automatically extract hierarchical features from raw traffic data, allowing increasingly complex spatial and temporal traffic patterns to be learned without explicit feature design~\cite{lecun2015deep}. 
Numerous neural network architectures have since been explored, including multilayer perceptrons, convolutional neural networks (CNNs), recurrent neural networks (RNNs), long short-term memory (LSTM) networks, autoencoders, graph neural networks (GNNs), and transformer-based models~\cite{ABBASI202119, dong2025deep}.
As these models have become increasingly complex, XAI has emerged as a complementary research direction for investigating and interpreting their decision-making processes~\cite{10.1145/3561048,e23010018, 10.1145/3583558}.

Recent work demonstrates that explainability provides value beyond increasing the transparency of black-box models. 
Garcia et al.~\cite{9887744} employed explanation techniques to understand model predictions, debug DL classifiers, and simplify model architectures. 
Nascita et al.~\cite{9490313} analyzed feature importance to assess trustworthiness and improve multimodal traffic classification models, while their subsequent work used explanations to investigate the behavior of incrementally trained classifiers under evolving traffic scenarios~\cite{10.1145/3630050.3630178}. 
Luis-Bisbé et al.~\cite{app14135466} applied GradCAM to reveal shortcut learning, data leakage, and limitations of existing evaluation protocols. 
Collectively, these works demonstrate that explainability can support domain-specific reasoning and guide model refinement through expert interpretation.

One of the early DL approaches to encrypted traffic classification is deep packet, proposed by Lotfollahi et al.~\cite{lotfollahi2020deep}. 
The authors introduced a one-dimensional CNN that automatically learns feature representations directly from raw packet bytes and demonstrated strong classification performance on the ISCX VPN-nonVPN benchmark dataset~\cite{draper2016characterization}. 
Building upon this architecture, our subsequent work~\cite{9941392} incorporated class activation maps (CAM), an explainability technique originally developed for computer vision, to visualize the packet regions contributing to individual predictions and thereby enable the inspection of the model's decision-making process~\cite{Zhou_2016_CVPR}.
The work adopted a design study methodology to elicit domain experts' explanations of data and model behavior, informing the iterative design of explanation visualizations aligned with established network analysis workflows and familiar analytical representations~\cite{6327248}. 
Building upon this work, the approach was further extended by supporting the transition from local explanation analysis to subset-based aggregation, allowing analysts to investigate the overall decision strategy learned by the model for specific application classes~\cite{10.1007/978-3-031-44067-0_1}.

Our preceding work investigated data drift and out-of-distribution detection~\cite{grivapp26}, two major challenges in deployed ML systems, where changes in the underlying data distribution may lead to degraded predictive performance over time~\cite{9829791}. 
To support the investigation of such distributional changes, the proposed VA approach combines XAI with interactive visual analysis, enabling domain experts to detect, interpret, and validate data shifts using their domain knowledge.

Collectively, these studies demonstrate that explainability alone is insufficient to support informed decision-making. 
Meaningful interpretation of both model behavior and data characteristics requires the active involvement of domain experts, whose expertise is essential for validating, contextualizing, and reasoning about the resulting insights. 
These observations motivated increased attention to sense-making, emphasizing the role of explanations in supporting expert reasoning and informed decision-making throughout the ML lifecycle~\cite{10.1145/3387166, 10.1145/3328485, hoffman2018metrics, 10.1145/3387166, miller2019explanation, 10.1145/3538882.3542790}.

Similar principles have long been established in the field of VA, where interactive visual interfaces are designed to combine automated analysis with human reasoning in order to facilitate knowledge generation~\cite{11033407, Keim2008}. 
Rather than replacing human expertise, VA frameworks explicitly model iterative processes of exploration, hypothesis generation, verification, and knowledge acquisition through close interaction between computational methods and domain experts.
Among these, the knowledge generation model proposed by Sacha et al.~\cite{6875967} provides a particularly suitable foundation, as it explicitly describes how human reasoning and computational analysis interact through iterative exploration and verification cycles to transform analytical findings into domain knowledge. These characteristics closely align with the objectives of explanation-based model analysis, where explanations serve not only to expose model behavior but also to support expert reasoning and informed decision-making.

Recent surveys on XAI for network traffic analysis report similar observations. 
Although existing approaches have demonstrated the value of explainability for understanding ML-based traffic classifiers, they also identify the need for more human-centered explanation strategies, improved support for different stakeholder groups, and tighter integration of explainability into the ML lifecycle~\cite{nascita2024survey}. 
These findings further reinforce the importance of combining explainability with interactive analysis and expert-driven sense-making.

\section{Limitations of Traffic Classification Workflows}



The proposed workflow is grounded in findings gathered throughout this research, including a systematic literature analysis, the development and evaluation of ML models for network traffic classification, the application of XAI techniques, empirical studies, expert interviews, and practitioner discussions~\cite{grivapp26, 10.1007/978-3-031-44067-0_1,9941392}. This evidence consistently reveals recurring shortcomings in benchmark datasets~(\ref{datasetlimitation}), current model evaluation practices~(\ref{performacecentriclimitation}), and the integration of explainability and expert knowledge into model development~(\ref{expertperspectivelimitations}), motivating the proposed workflow.

\subsection{Dataset-Related Limitations} \label{datasetlimitation}

A recurring observation across network traffic classification workflows is that model development is often emphasized more strongly than systematic analysis of the underlying datasets.
While many studies remove obvious identifiers such as ethernet headers or source and destination IP addresses to reduce information leakage, less apparent sources of shortcut information often remain unnoticed.
They may only become apparent through detailed dataset inspection or by analyzing model behavior using XAI techniques. 
Such analyses provide insights into the model's decision-making process and reveal whether predictions are driven by meaningful traffic characteristics or unintended shortcut features.

To investigate this issue, we analyzed the ISCX VPN-nonVPN dataset, one of the widely adopted benchmark datasets for encrypted network traffic classification~\cite{SHARMA2025110984, HAN2026112524}. 
The dataset was constructed by generating representative real-world network activities across a diverse set of applications, including web browsing, streaming, VoIP, chat, file transfer, email, and peer-to-peer communication~\cite{draper2016characterization}. 
For each application scenario, traffic was captured both as regular network traffic and while routed through a VPN, resulting in fourteen traffic categories that represent a broad spectrum of encrypted communication patterns.

Data-induced artifacts represent an often overlooked source of bias in network traffic classification datasets. 
Such artifacts unintentionally correlate with the target classes and may enable ML models to exploit shortcut features instead of learning meaningful characteristics of network traffic. 
To better understand the presence of such artifacts and account for them during the design of future benchmark datasets, we examine specific structural properties of the dataset from an artifact-oriented perspective.

A network flow is a fundamental abstraction in network traffic analysis that represents a sequence of packets belonging to the same communication session between two communicating endpoints.
For the purpose of our analysis, each flow is represented solely by its transport-layer identifier, i.e., the tuple (source port, destination port, protocol), deliberately excluding IP addresses and TCP sequence numbers. 
One potential source of data-induced artifacts arises from transport-layer flow identifiers that occur exclusively within a single application. 
Such application-specific flow identifiers may unintentionally encode the target class and therefore enable shortcut learning. 
To quantify this effect, we measure the fraction of unique flows, i.e., flow identifiers occurring exclusively within a single application, and the fraction of unambiguous packets, i.e., packets belonging to such unique flows (Table \ref{tab:flow-separability}). 
Together, these metrics assess the extent to which transport-layer identifiers alone distinguish application classes.
\begin{table}[t]
  \centering
  \footnotesize 
  \caption{Flow-level separability of the dataset. Unique flows is the
    fraction of an application's flow identifiers
    (src port, dst port, protocol) that appear in no other application;
    Unambiguous packets is the fraction of its packets carried by those
    unique flows. 
    High values indicate that the class is trivially separable
    from the flow identifier alone, before any payload inspection.}
  \label{tab:flow-separability}
  \begin{tabular}{l S[table-format=5.0] S[table-format=7.0] S[table-format=1.2] S[table-format=1.2]}
    \toprule
    {Application} & {Number of flows} & {Number of packets} & {Unique flows} & {Unambiguous packets} \\
    \midrule
    AIM chat   & 399   & 4974    & 0.64 & 0.49 \\
    Email      & 3057  & 51098   & 0.84 & 0.71 \\
    Facebook   & 1463  & 556912  & 0.87 & 0.98 \\
    FTPS       & 551   & 2220823 & 0.97 & 0.99 \\
    Hangouts   & 1466  & 1869579 & 0.87 & 0.99 \\
    ICQ        & 404   & 7476    & 0.66 & 0.56 \\
    Netflix    & 286   & 732836  & 0.88 & 0.23 \\
    SCP        & 26    & 183251  & 0.96 & 0.99 \\
    SFTP       & 144   & 423560  & 0.91 & 0.99 \\
    Skype      & 4671  & 1470613 & 0.88 & 0.98 \\
    Spotify    & 299   & 97470   & 0.93 & 0.25 \\
    Torrent    & 629   & 269096  & 0.98 & 0.98 \\
    VoipBuster & 2436  & 1459926 & 0.78 & 0.99 \\
    Vimeo      & 551   & 366694  & 0.91 & 0.86 \\
    Youtube    & 803   & 268310  & 0.96 & 0.69 \\
    \hdashline
    {Average} & 1146 & 665508 & 0.87 & 0.78 \\
    \bottomrule
  \end{tabular}
\end{table}
The analysis reveals that transport-layer identifiers alone provide substantial discriminative information. 
On average, 87\% of all flow identifiers are unique to a single application, while 78\% of all packets belong to such unique flows. For several applications, including Torrent, FTPS, SCP and SFTP, nearly all traffic can be distinguished solely by the transport-layer identifier before considering any payload. 
Consequently, a model can achieve high predictive performance by implicitly learning application-specific port and protocol combinations rather than meaningful characteristics contained within the application-layer data carried in the packet payload.

\begin{figure}[htbp]
    \centering
    \includegraphics[width=1.0\textwidth]{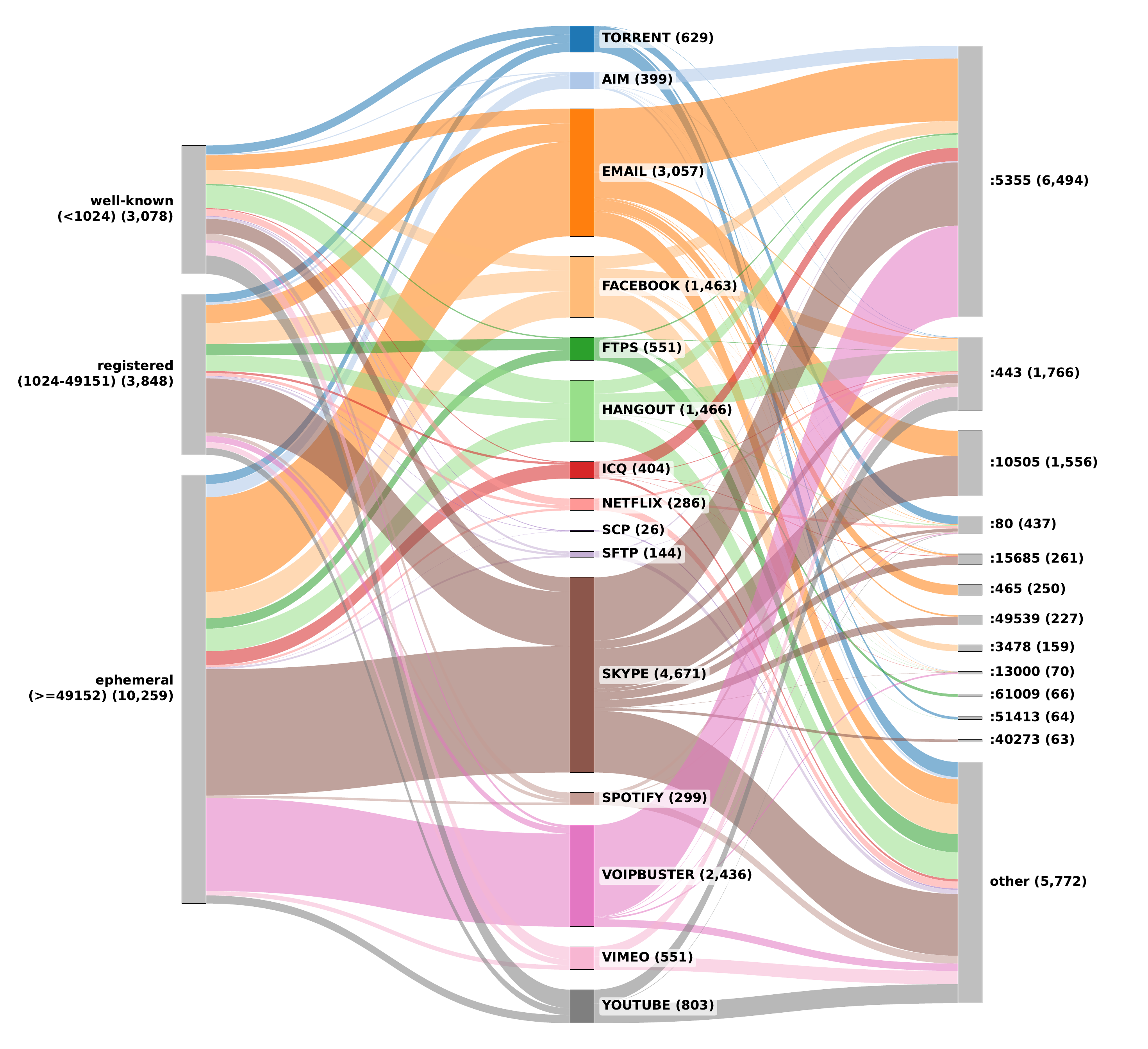}
\caption{Sankey diagram of the per-application flow structure. Each flow, characterized by its signature (\texttt{src\_port}, \texttt{dst\_port}, \texttt{protocol}), is traced through three stages: source-port band (left), application class (centre), and destination port (right). Ribbon width is proportional to the number of distinct flow signatures and is coloured by application; source ports are grouped into the three IANA ranges, and the twelve most frequent destination ports are shown individually while the remainder are aggregated as \emph{other}. Node labels give the number of flows.}
    \label{fig:sankey}
\end{figure}
The analysis further demonstrates that shortcut behavior is highly application-dependent. For applications such as Netflix and Spotify, most flow identifiers remain unique, whereas the majority of packets are transmitted through a small number of shared, high-volume flows. Consequently, flow-level and packet-level separability differ substantially, indicating that the structural properties of individual application classes can strongly influence the composition of the training and test sets. 
This observation is particularly relevant because many existing studies use random packet-level train-test splits. As a result, packets from the same network flow can appear in both the training and test sets. Since packets within a flow share common communication characteristics, this introduces information leakage and may lead to overly optimistic estimates of classification performance.

Figure~\ref{fig:sankey} provides a holistic view of the flow structure of the ISCX VPN-nonVPN dataset by visualizing every unique flow signature as a three-stage Sankey diagram. 
Each flow is represented by its transport-layer identifier, i.e., the tuple (src\_port, dst\_port, protocol), connecting the source-port range, the corresponding application class, and the destination port. 
Source ports are grouped according to the three IANA ranges (well-known, registered, and ephemeral), while the most frequent destination ports are shown individually and less frequent ports are aggregated into an other category. 
Ribbon width corresponds to the number of distinct flow signatures, enabling the structural relationships between transport-layer identifiers and application classes to be examined at a glance.
The visualization reveals that the flow structure is highly application-dependent and far from uniformly distributed. 
Although most flows originate from ephemeral source ports, substantial numbers also originate from well-known and registered ports. 
On the destination side, the distribution is dominated by a small number of ports, most notably ports 5355 and 443, while many remaining ports occur only rarely. 
More importantly, several destination ports are shared across multiple applications, whereas others are almost exclusively associated with a single application.
Applications whose transport-layer identifiers predominantly map to unique destination ports can be distinguished largely from their flow identifiers alone, whereas applications sharing common ports require additional discriminative information. 
Consequently, the intrinsic classification difficulty varies considerably between application classes before any packet payload is analyzed. 
These observations complement the quantitative results presented in Table~\ref{tab:flow-separability} and demonstrate that the dataset itself already contains structural properties capable of biasing model learning toward transport-layer artifacts rather than characteristics of the encrypted traffic.

Beyond transport-layer identifiers, additional shortcut sources may arise from the data collection process itself. 
For example, if traffic belonging to a particular application is captured during a specific time period or under consistent experimental conditions, temporal or environmental characteristics may become correlated with the target class, creating sampling-induced shortcuts that are unrelated to the actual application behavior. 
Such correlations can originate from the order of data collection, network configuration, operating system state, background traffic, or other experimental factors that are unintentionally preserved in the dataset. 
These artifacts are often difficult to identify through conventional preprocessing alone and typically require systematic dataset analysis or explanation-based inspection to become apparent. 
The observations presented are further supported by insights gathered through interviews and discussions with domain experts conducted during our evaluations, and through conversations with network analysis practitioners at conferences such as SharkFest.


\subsection{Limitations of Performance-Centric Evaluation} \label{performacecentriclimitation}
\begin{table}[t]
\centering
\caption{Overall performance (\%, mean $\pm$ std). Packets are represented by the first 600 bytes from the IP header (TCP/UDP payload packets only). The CNN is initially trained with source and destination IP addresses masked. Tests~1--3 evaluate the same model under progressively stronger masking at test time (IP $\rightarrow$ IP+ports $\rightarrow$ IP+ports+TCP sequence/acknowledgement). The retrained model is trained and evaluated with all four fields masked. CNN results are averaged over 5 random seeds;
As a complementary experiment, we trained a HistGradientBoostingClassifier (HGB) over 10 seeds using only the source port, destination port, and TCP sequence number (set to zero for UDP).
}
\label{tab:overall}
\begin{tabular}{llccc}
\toprule
Setting & Test-time masking & Accuracy & Macro-$F_1$  \\
\midrule
CNN, Test 1 (train mask \texttt{\{ip\}})   & \texttt{\{ip\}}              & $93.62 \pm 0.06$ & $93.71 \pm 0.06$ &  \\
CNN, Test 2 (train mask \texttt{\{ip\}})   & \texttt{\{ip,seq,ack\}}      & $87.94 \pm 1.96$ & $87.92 \pm 2.18$ &  \\
CNN, Test 3 (train mask \texttt{\{ip\}})   & \texttt{\{ip,seq,ack,port\}} & $50.21 \pm 1.58$ & $45.86 \pm 1.92$ & \\
CNN, retrained masked                      & \texttt{\{ip,seq,ack,port\}} & $91.12 \pm 0.15$ & $91.20 \pm 0.15$ & \\
\midrule
Tabular (HGB, ports+seq only)              & ---                          & $93.28 \pm 0.17$ & $93.57 \pm 0.16$\\
\bottomrule
\end{tabular}
\end{table}

The performance and trustworthiness of ML models are inherently constrained by the quality of the data used for training. 
Regardless of the complexity of the learning algorithm, models can only learn from the information present in the dataset. 
Consequently, biases, artifacts, and unintended correlations introduced during data collection or preprocessing may become part of the learned decision strategy and remain unnoticed during conventional model evaluation. 
As demonstrated in the previous subsection~\ref{datasetlimitation}, these representations may still contain capture metadata, lower-layer protocol information, and other dataset-specific artifacts that unintentionally encode class labels, thereby allowing models to exploit shortcut features instead of learning meaningful characteristics of encrypted traffic.
Table~\ref{tab:overall} illustrates the impact of transport-layer identifiers on model performance. While progressively masking these fields leads to a substantial degradation in CNN accuracy, the identifier-only baseline achieves nearly the same performance using only source port, destination port, and TCP sequence number. Together, these results indicate that much of the predictive performance is driven by shortcut features rather than application-specific characteristics.

The choice of evaluation metric is equally important, particularly for the highly imbalanced datasets commonly encountered in network traffic classification. 
For this reason, most studies appropriately report the macro F1-score, which weights all classes equally and therefore provides a more balanced assessment of classification performance~\cite{10858503, 11023272}. 
However, the reported F1-score is often insufficiently specified. 
Many studies simply state that precision, recall, and F1-score are computed from the numbers of true positives (TP), false positives (FP), and false negatives (FN), without indicating whether the reported F1-score corresponds to the micro, macro, or weighted average in the multiclass setting. 
This ambiguity complicates the interpretation and comparison of reported results. 
The use of the micro F1-score is dominated by majority classes and may therefore conceal poor performance on underrepresented applications~\cite{zhao2023yet, wang2024netmamba}.
Aceto et al. additionally report the G-mean, which combines sensitivity and specificity to provide a more balanced evaluation under class imbalance~\cite{8640262}.
Some studies construct artificially balanced datasets by sampling an equal number of instances per class. 
In these circumstances, metric such as accuracy are appropriate.
Beyond aggregated performance metrics, confusion matrices are widely used as an intuitive visual tool for analyzing class-specific prediction behavior, making systematic misclassifications and confusion between individual classes readily identifiable.

Predictive performance alone provides only a limited assessment of model quality. 
High classification performance does not necessarily indicate that a model has learned meaningful characteristics of network traffic. 
Rather, performance metrics alone cannot determine whether predictions are based on semantically meaningful traffic patterns or on the exploitation of dataset-specific shortcut features.




\subsection{Expert Perspectives and Explainability} \label{expertperspectivelimitations}

High predictive performance alone is insufficient for the practical deployment of classification models. 
Beyond achieving competitive benchmark results, practitioners require transparent and interpretable models that enable them to understand, verify, and validate model predictions before they can be trusted in operational environments. 
Identified shortcomings may indicate the need for improved dataset construction, revised preprocessing strategies, alternative model architectures, or modified evaluation procedures.
This observation is consistent with recent literature advocating explainability as a prerequisite for trustworthy ML and was further reinforced by our own experiences during model evaluation, interviews with domain experts, and discussions with network analysis practitioners at technical conferences.
Consequently, expert-guided interpretation should not be viewed as an isolated post-hoc analysis but rather as an iterative process that supports continuous improvement throughout the entire model development and deployment lifecycle.

XAI provides an important mechanism for addressing this gap. 
By exposing the reasoning underlying model predictions, XAI enables experts to validate learned decision patterns against established domain knowledge, identify misleading or spurious features, and assess whether the model relies on semantically meaningful traffic characteristics. 
This not only increases confidence in model predictions but also facilitates a deeper understanding of model limitations and failure cases that remain hidden when relying solely on predictive performance metrics.


The integration of XAI benefits from an iterative design study process~\cite{6327248}.
Such a process combines the development of visual explanation techniques, continuous collaboration with domain experts using their familiar analytical tools, and repeated cycles of validation and reflection. 
Throughout these iterations, explainability evolves from a post-hoc interpretation tool into an integral component of model validation, dataset inspection, and iterative model improvement. 
In addition, it enables the extraction of domain insights and strengthens experts' trust in both the learned models and their explanations.
A first observation is that explanations inherently require context. 
While explanation methods can identify influential bytes, protocol fields, or packet regions, these highlighted features have little meaning in isolation. 
Their relevance can only be assessed by relating them to protocol semantics, application behavior, and networking expertise. 
Consequently, explanations cannot be interpreted automatically but require human reasoning to determine whether influential features correspond to meaningful characteristics of traffic or merely reflect implementation details, protocol structures, or unintended dataset artifacts.
Both expert studies emphasize that explainability acts primarily as an interface between machine learning developers and network analysts~\cite{grivapp26,9941392}.

\begin{figure}[htbp]
    \centering
    \begin{minipage}{0.48\textwidth}
        \centering
        \includegraphics[width=\linewidth]{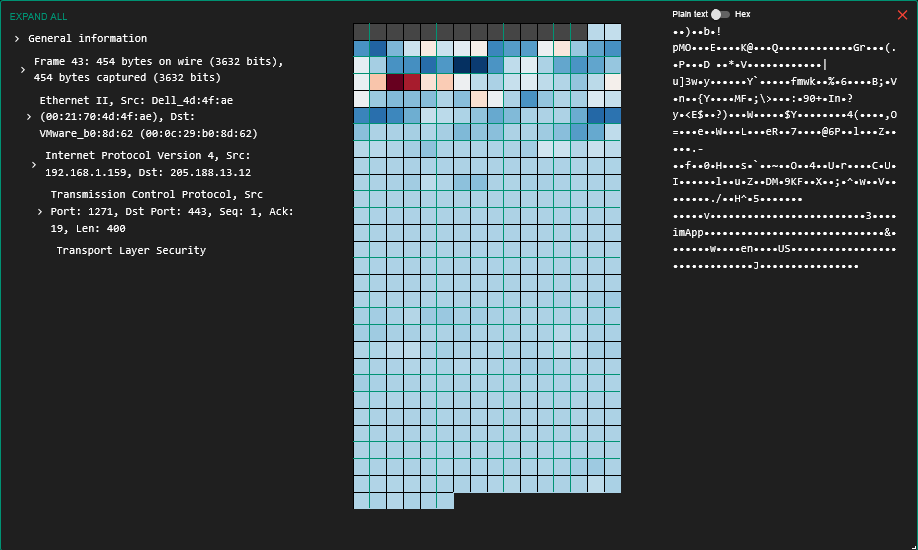}
        \caption{Local explanation of an individual network traffic sample. The attribution map highlights the contribution of each input byte to the model's prediction, allowing experts to inspect the decision rationale for a single classification~\cite{9941392}.}
        \label{fig:local}
    \end{minipage}
    \hfill
    \begin{minipage}{0.48\textwidth}
        \centering
        \includegraphics[width=\linewidth]{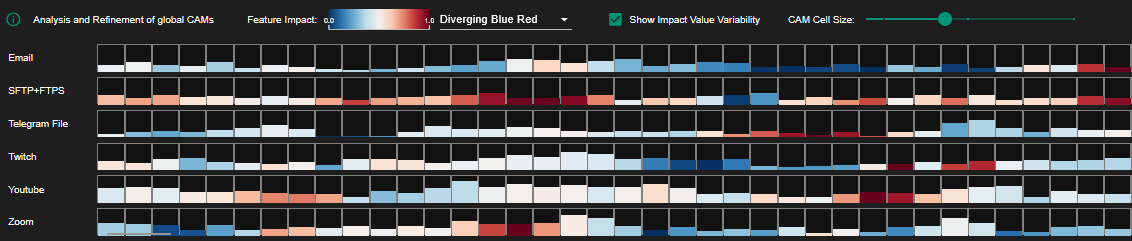}
        \caption{Global explanations of application classes obtained by aggregating local attribution maps across multiple samples. The visualization enables experts to identify consistent attribution patterns at the class level~\cite{cherepanov2026interactiveanalysisglobalexplanations}.}
        \label{fig:global}
    \end{minipage}
\end{figure}

Applying XAI revealed that only the initial bytes of the encrypted payload consistently contribute to the model's predictions, whereas the majority of the payload receives negligible attribution scores. 
This behavior is illustrated in Figures~\ref{fig:local} and~\ref{fig:global}, which show the attribution values for an exemplary packet classification and the corresponding aggregated attribution values at the application-class level, respectively.
This indicates that large portions of the encrypted payload should not contain discriminative information for the classification task, since correctly encrypted payload bytes are intended to be indistinguishable from random noise.
Consequently, substantially smaller input representations may be sufficient, reducing computational complexity while maintaining predictive performance. 
This observation is supported by previous work \cite{grivapp26}, which demonstrated that restricting the model input to the informative payload region does not degrade classification performance.
The interpretation of such findings requires confirmation by domain experts before they are incorporated into feature selection, preprocessing, or model refinement.

Another important lesson concerns the aggregation of explanations. 
While local explanations provide insights into individual predictions, practitioners are often interested in understanding the overall decision strategy learned for a subset of data (e.g., an application class, illustrated in Figure~\ref{fig:global}). 
This requires representative global explanations.
Numerous mathematical aggregation strategies exist, each emphasizing different properties of the underlying explanation values. 
Selecting an appropriate aggregation strategy is consequently essential for obtaining meaningful global explanations. 
Moreover, the resulting explanations require expert interpretation and should support interactive exploration of different data subsets. 
Besides selecting an appropriate aggregation strategy, particular attention must be paid to the intuitive visual representation of global explanations, as aggregating explanations from many samples introduces additional visualization and interpretation challenges than explaining individual predictions.

Collectively, these findings motivate the need for a systematic workflow that supports explanation-based model validation and semantic assessment. Section~\ref{four} introduces the proposed workflow.

\section{Specific Workflow for Network Traffic Classification} \label{four}
\begin{figure}[htbp]
    \centering
    \includegraphics[width=0.8\textwidth]{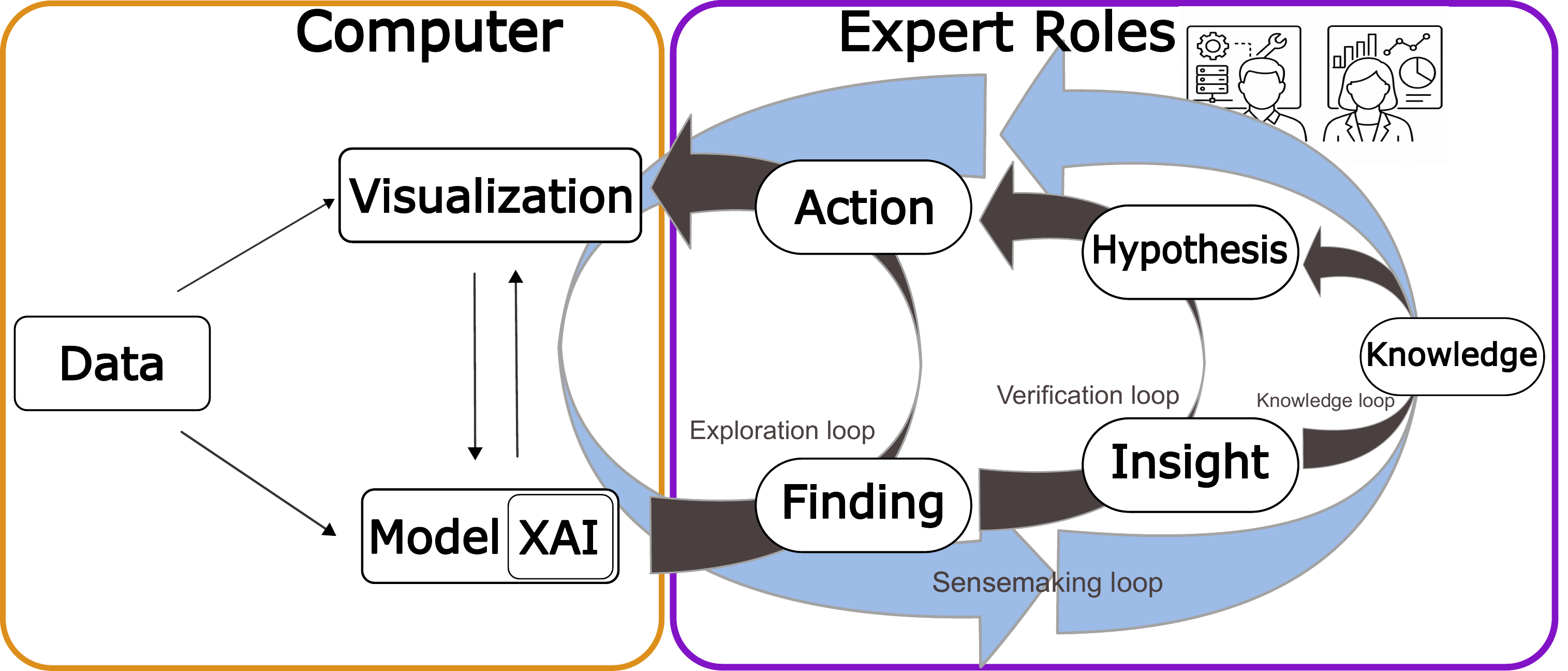}
\caption{Adaptation of the knowledge generation model for VA to network traffic classification. Compared with the original model~\cite{6875967}, the framework explicitly incorporates XAI into the computational pipeline and refines the human component into multiple expert roles with stakeholder-specific objectives, supporting explanation-based exploration, verification, and knowledge generation}
    \label{fig:framework}
\end{figure}

Motivated by the findings presented in the previous section, we apply the knowledge generation model for VA proposed by Sacha et al.~\cite{6875967, SACHA2017164} to support the development and semantic validation of network traffic classification models.
This model provides the most suitable conceptual foundation for our work because it explicitly integrates computational analysis with human cognitive reasoning. 
Existing process models differ in their objectives. 
Data science methodologies, such as KDD~\cite{Fayyad199627}, CRISP-DM~\cite{Chapman2000CRISPDM1S}, and CRISP-ML(Q)~\cite{make3020020}, focus on the engineering lifecycle of ML systems, including data preparation, model development, evaluation, and deployment. 
They primarily assess models through predictive performance and operational quality, without incorporating XAI, interactive visual analytics, or expert-driven semantic assessment. 
Visualization and visual analytics models, such as those proposed by van Wijk~\cite{1532781}, and Keim et al.~\cite{Keim2008}, move closer to our objective by integrating data, analytical models, visualization, and user interaction to support knowledge generation. 
Nevertheless, these frameworks do not explicitly represent the human reasoning process and support the semantic assessment of ML models.

The knowledge generation model proposed by Sacha et al. consists of two tightly coupled components, as illustrated in Figure~\ref{fig:framework}. The computer system comprises network traffic data, ML models, and interactive visualizations.
The human component represents the analyst's cognitive processes, including interpretation, reasoning, hypothesis generation, and decision-making based on visualized data, model information, and classification explanations.
In the following, we describe how these components are instantiated and adapted to the domain of network traffic classification.

\subsection{Data}

The data component comprises the structured network traffic data used for model training and assessment. 
Since the framework addresses application classification, the dataset is assumed to be labeled. 
Although the data follows a structured representation, its characteristics depend on the underlying network protocols, each providing protocol-specific fields and semantics.

From a functional perspective, the data must support appropriate preprocessing and transformation steps prior to model development. 
This includes the identification and removal or masking of known sources of bias and dataset-specific artifacts that could lead to shortcut learning rather than meaningful feature extraction. 
Common examples include masking IP addresses, port numbers, sequence and acknowledgment numbers, and removing Ethernet frame information.
Padding or truncation may also be required as part of the preprocessing pipeline to obtain a homogeneous input representation for the model. 

In addition, the data should satisfy several non-functional requirements. 
It should accurately reflect real-world application traffic to ensure meaningful semantic evaluation, be sufficiently complete with minimal missing or corrupted values, and provide representative coverage of the application classes. These properties are essential to ensure that the generated explanations correspond to genuine traffic characteristics rather than artifacts of the dataset itself.

The data partitioning strategy should preserve the independence of the training, validation, and test sets. 
All packets belonging to the same flow must be placed exclusively in either the training, validation, or test set and must never be distributed across multiple partitions. Otherwise, the model may exploit flow-specific characteristics shared between packets of the same flow, leading to information leakage, overly optimistic performance estimates, and an unreliable assessment of its generalization capability.
\subsection{Visualization}

\begin{table*}[t]
\centering
\caption{Stakeholder-specific objectives and explanation requirements.}
\label{tab:expertroles}
\scriptsize
\begin{tabularx}{\textwidth}{p{3cm}XX}
\toprule
\textbf{Aspect} &
\textbf{ML Experts} &
\textbf{Network Experts} \\
\midrule

Primary objective &
Understand, validate, and improve the behavior of the ML model. &
Validate whether the learned decision strategies correspond to meaningful network patterns. \\

Expert knowledge &
ML optimization, model architectures, explainability methods, feature engineering. &
Network protocols, packet structures, traffic analysis, application behavior, communication semantics. \\

Typical analytical tasks &
Model debugging, shortcut detection, robustness analysis, explanation validation, comparison of alternative models. &
Protocol validation, semantic interpretation, artifact detection, deployment assessment, traffic investigation. \\

Primary evaluation focus &
Faithfulness, robustness, consistency, sensitivity, and explanation reliability. &
Semantic correctness, usefulness and transferability. \\

Knowledge generated &
Knowledge about model behavior, explanation reliability, and opportunities for improving the learning process. &
Knowledge about semantic validity, protocol behavior, dataset quality, and deployment suitability. \\

Feedback to the workflow &
Model refinement, preprocessing improvements, feature engineering, training strategy. &
Dataset refinement, semantic validation, feature engineering, data collection, deployment decisions. \\
\bottomrule
\end{tabularx}
\end{table*}

The visualization component represents the interface through which analytical results are transformed into actionable knowledge. 
Rather than serving solely as a means of presenting data, visualizations should support the exploration, interpretation, and validation of both the underlying data and the generated explanations. 
Consequently, visualization acts as the primary communication channel between automated analysis and human reasoning.

From a functional perspective, visualizations should present information in a manner that is intuitive and consistent with the analytical task. 
Since explanations can be provided at both local and global levels, the visualization should support the exploration of individual traffic instances as well as aggregated explanations describing application classes or the classifier as a whole.

Different user groups interact with XAI systems with fundamentally different objectives~(Table \ref{tab:expertroles}), which directly influence their explanation requirements. 
ML practitioners primarily seek explanations to debug and improve models, identify erroneous or biased decision behavior, understand which information the model utilizes, and assess its strengths and limitations. 
In contrast, domain experts are typically interested in understanding and validating model predictions within the application context, integrating model outputs into downstream decision-making, learning new domain knowledge, assessing prediction reliability, or contesting model decisions when necessary. 
Consequently, the visualization should not adopt a one-size-fits-all approach but instead provide user-centered explanation interfaces that adapt the presented information and level of detail to the analytical goals and expertise of the intended users (context). 
Moreover, explanations should remain coherent with networking concepts and protocol semantics to support meaningful interpretation~\cite{10316181} (Table~\ref{tab:xai_evaluation}).

\begin{table}[htbp]
\centering
\scriptsize
\caption{Presentation- and user-related evaluation requirements for XAI explanations in network traffic classification~\cite{10.1145/3583558}.}
\label{tab:xai_evaluation}
\begin{tabular}{p{1.8cm} p{2.cm} p{8.cm}}
\hline
\textbf{Category} & \textbf{Characteristic} & \textbf{Objective} \\
\hline
Presentation
& Compactness
& Explanations should remain concise by highlighting only the most relevant features and avoiding redundant information. \\
& Composition
& Explanations should be presented using intuitive and well-structured visual representations that are appropriate for the explanation type (e.g., local or global) and support efficient analytical workflows. \\

\hline

\multirow{3}{*}{User}
& Context
& Explanations should provide information relevant to the analytical objectives and domain knowledge of network analysts. \\

& Coherence
& Explanations should be consistent with networking knowledge and protocol semantics while avoiding dataset-specific artifacts. \\

& Controllability
& The explanation interface should support interactive exploration, comparison of traffic subsets, and investigation of alternative scenarios. \\

\hline
\end{tabular}
\end{table}

Furthermore, visualizations should facilitate interactive exploration by allowing analysts to inspect different subsets of traffic, compare explanations across applications or models, and investigate alternative scenarios (see controllability in Table~\ref{tab:xai_evaluation}). 
Such interaction enables analysts to iteratively formulate and validate hypotheses, thereby supporting the knowledge generation process rather than merely presenting static results.

Finally, the visualization should minimize cognitive effort by emphasizing relevant information while avoiding unnecessary visual complexity. Explanation representations should remain compact.
Their composition should employ intuitive, well-structured visual representations that are appropriate for the explanation type.
Furthermore, explanations should provide sufficient context by relating highlighted features to the analyst's objectives and domain knowledge, while maintaining coherence with networking concepts and protocol semantics (Table~\ref{tab:xai_evaluation}). 

\subsection{Model}
The model component represents the ML model responsible for learning patterns from the input data and correctly classifying network traffic into the corresponding classes. 
Its primary goal is to learn meaningful decision patterns that generalize beyond the training data rather than memorizing dataset-specific characteristics.

Besides achieving high predictive performance, the model should demonstrate good generalization to unseen data and robustness against variations in the input. Accordingly, model evaluation should include cross-validation to obtain reliable performance estimates, multiple training runs with different random initializations to assess stability, and robustness testing on unseen or perturbed data to evaluate the reliability and generalizability of the learned decision function~\cite{LONES2024101046}.

The model should be evaluated using metrics appropriate for the underlying class distribution. 
For imbalanced datasets, metrics such as macro F1-score, precision, and recall provide a more informative assessment than accuracy alone~\cite{SOKOLOVA2009427}. 
For balanced datasets, accuracy is also an appropriate performance measure. 
In addition, confusion matrices provide valuable insights into the model's prediction behavior by revealing application classes that are frequently confused and therefore require further investigation.

\subsection{XAI Methods}
XAI comprises a broad range of methods that provide insights into the decision-making process of machine learning models. Existing approaches can be broadly categorized into intrinsic and post-hoc methods, with the latter being most commonly applied in network traffic classification because they can explain already trained black-box models without modifying their architecture~\cite{BARREDOARRIETA202082}. 
Post-hoc methods can further be distinguished into feature attribution techniques, concept-based explanations, surrogate models, example-based explanations, and counterfactual explanations~\cite{10.1145/3583558,10.1145/3561048}. Feature attribution methods, such as SHAP, LIME, Integrated Gradients, Layer-wise relevance propagation (LRP), and CAM, are among the most widely used approaches because they identify the input features that contribute most strongly to individual predictions~\cite{10.1145/3561048, BARREDOARRIETA202082}.

\begin{table*}[t]
\centering
\caption{Content-related evaluation requirements for XAI explanations in network traffic classification~\cite{10.1145/3583558}.}
\label{tab:xai_content}
\scriptsize
\begin{tabularx}{\textwidth}{p{1.4cm}p{2cm}Xp{4.5cm}}
\toprule
\textbf{Category} &
\textbf{Characteristic} &
\textbf{Objective} &
\textbf{Example Evaluation} \\
\midrule

\multirow{3}{*}{Content}
& Correctness
& Explanations should faithfully represent the reasoning of the underlying classifier rather than producing only plausible feature attributions.
& Feature deletion/insertion, perturbation analysis. \\

& Consistency
& Identical traffic samples should produce identical explanations independent of implementation details.
& Repeated explanations for identical inputs, invariance across equivalent models, consistency across different random initializations\\

& Continuity
& Similar network flows or packets should produce similar explanations, demonstrating robustness to small variations in the input.
& Stability under slight input perturbations, explanation similarity for neighboring flows/packets, fidelity under small traffic variations. \\

\bottomrule
\end{tabularx}
\end{table*}

Because explanations serve as the basis for semantic model validation and expert reasoning, their quality must also be assessed. Therefore, we incorporate an XAI evaluation component into the framework (Figure~\ref{fig:framework}). This component provides a structured approach for assessing the quality of explanations generated for network traffic classification. Unlike predictive performance, explanation quality cannot be adequately characterized by a single metric. 
The evaluation requirements adopted in this framework are summarized in Tables~\ref{tab:xai_content} and are adapted from established explanation quality dimensions proposed in the XAI literature~\cite{10.1145/3583558}. 
From the content-related characteristics, we focus on correctness, consistency, and continuity because they evaluate whether explanations faithfully represent the model's reasoning.

The content-related characteristics are primarily evaluated through technical, functionally grounded evaluation methods that assess the behavior of the explanation algorithm independently of human judgment. 
Examples include perturbation analyses, feature deletion and insertion experiments, repeated explanation generation, and robustness analyses under small input variations. 

\subsection{Exploration Loop}

The exploration loop describes how analysts iteratively investigate explanation results to identify meaningful patterns in model behavior and network traffic. Exploration may begin with a suspicious prediction, an unexpected feature attribution, or a hypothesis regarding protocol-specific behavior. Analysts can inspect local explanations for individual traffic instances, aggregate explanations across traffic subsets or application classes, compare different models, and switch between local and global perspectives to identify recurring explanation patterns and potential anomalies.

The resulting findings provide insights into the classifier's decision-making process and may reveal discriminative traffic characteristics, protocol-specific behaviors, unexpected feature attributions, shortcut learning, or dataset-specific artifacts. 
These findings represent candidate explanations of the observed model behavior and form the basis for subsequent verification.

\subsection{Verification Loop}

The verification loop systematically evaluates candidate findings identified during exploration to determine whether they reflect semantically meaningful classifier behavior or originate from shortcut learning, dataset artifacts, or limitations of the explanation method. Verification should therefore combine complementary sources of evidence, including alternative explanation methods, quantitative evaluation metrics, additional datasets, different model architectures, and expert assessment.

Depending on the investigated hypothesis, verification may examine whether highlighted protocol fields correspond to known communication behavior, whether similar explanation patterns are consistently observed across different explanation methods, models, or datasets, and whether domain experts consider the identified decision strategy plausible and consistent with protocol semantics. Quantitative analyses provide evidence regarding the reproducibility and robustness of the observed explanation patterns, while expert interpretation determines whether these patterns are meaningful from a networking perspective.

Only findings that are consistently supported by complementary technical evidence and expert interpretation should be regarded as validated knowledge. Such knowledge can subsequently guide refinements of the dataset, feature representation, explanation strategy, or the underlying ML model, thereby supporting the development of more robust and semantically meaningful traffic classifiers.
For example, if technical analyses consistently show that a classifier relies on the destination port to identify encrypted Spotify traffic, a network expert may conclude that this behavior reflects a dataset-specific artifact rather than semantically meaningful traffic characteristics. This insight may motivate masking the destination port or redesigning the feature representation before retraining the model.

\subsection{Knowledge Generation Loop}

The knowledge generation loop consolidates validated findings into reusable knowledge that can support future model development and evaluation. Rather than focusing on individual explanations or specific observations, it captures general insights about network traffic characteristics, model behavior, explanation methods, and the overall development process.

The generated knowledge may describe reliable protocol-specific communication patterns, discriminative traffic characteristics, common sources of shortcut learning, dataset limitations, or recurring explanation patterns observed across different models and datasets. It may also identify effective explanation techniques, visualization strategies, or evaluation procedures that consistently support semantic model validation.

Unlike the verification loop, which determines whether individual findings are supported by sufficient evidence, the knowledge generation loop generalizes these validated findings into reusable insights and recommendations. 

The resulting knowledge can subsequently guide future dataset construction, feature engineering, explanation design, visualization development, model evaluation, and ML model development. As this knowledge is reused in subsequent analyses, it influences new analytical objectives, supports more efficient exploration, and contributes to the continuous improvement of explainable network traffic classification workflows.

To facilitate the practical application of the proposed framework, Table~\ref{tab:recommendations} summarizes the main design guidelines associated with each framework component.
\begin{table}
\caption{Design recommendations for trustworthy network traffic classification derived from the adapted knowledge generation framework}
\label{tab:recommendations}
\scriptsize
\begin{tabularx}{\linewidth}{p{2cm}p{4.2cm}X}
\toprule
\textbf{Guideline category} & \textbf{Recommendation} & \textbf{Purpose} \\
\midrule

Data & Remove artifacts and ensure independent data partitions &
Reduces shortcut learning and information leakage. \\

Model &
Evaluate predictive performance using appropriate metrics and assess robustness, stability, and generalization. &
Provides evidence of the model's predictive performance and reliability. \\

XAI &
Evaluate explanation correctness, consistency, continuity, and comprehensibility &
Ensures that explanations can be reliably interpreted for semantic model analysis. \\

Visualization & Design interactive visualizations tailored to stakeholder-specific explanation requirements &
Supports efficient exploration and interpretation of explanation results.  \\

Exploration &
Explore explanation patterns across traffic subsets, application classes, and model variants. &
Identifies candidate findings, anomalies, and potential shortcut learning.  \\

Verification &
Verify candidate findings using complementary technical evidence and expert assessment &
Determines whether observed patterns are reproducible and semantically meaningful. \\

Knowledge &
Capture validated findings as reusable design recommendations and best practices. &
Supports future dataset design, model development, explanation design, and evaluation. \\
\bottomrule
\end{tabularx}
\end{table}

\section{Discussion and Future Work}

The proposed framework provides a structured mechanism for capturing knowledge generated during explanation-based model analysis and feeding it back into earlier stages of the development process. Beyond supporting the analysis of individual models, this enables recurring findings, such as protocol-specific shortcut features, dataset artifacts, explanation patterns, and effective preprocessing strategies, to be systematically documented and reused across subsequent studies. Such accumulated knowledge may improve the design of future benchmark datasets by increasing their transparency, documenting known sources of information leakage, and motivating standardized preprocessing and partitioning strategies. 
It also improves the reproducibility of analyses by explicitly documenting the rationale behind dataset modifications and model design decisions.

Because the proposed framework separates explanation generation from explanation analysis, it provides a common evaluation process in which different XAI methods can be assessed under consistent evaluation criteria and analytical tasks. This enables future work to compare explanation techniques not only with respect to their technical characteristics but also according to their practical utility during semantic validation. Furthermore, it facilitates the investigation of explanation-specific design choices, such as how local attribution values should be aggregated into representative global explanations and which aggregation strategies are most suitable for different analytical objectives.
For example, aggregating explanations across an entire application class or only across the most similar samples within a class.

A limitation of the proposed framework is the additional effort required to integrate explanation generation, interactive analysis, and expert-driven verification throughout the ML pipeline. Depending on the application scenario, incorporating all workflow components may not always be feasible because of constraints in time, computational resources, or domain expertise. Nevertheless, the framework is designed to support incremental adoption. Even the integration of individual components establishes a consistent process for documenting findings and feeding validated insights back into subsequent development iterations. As additional workflow components are incorporated, the accumulated knowledge base expands, enabling the framework to be continuously refined through practical application and facilitating the transfer of validated insights across different datasets, models, and traffic classification scenarios.

\section{Conclusion}
This paper addressed the limitations of performance-centric evaluation in network traffic classification by emphasizing the importance of semantic model validation. Building on findings synthesized from the literature, empirical analyses, practical experience with explainable traffic classification, and expert feedback, we identified key considerations for integrating explanation-based analysis into the model development process. These findings motivated the adaptation of the knowledge generation model for VA to the domain of network traffic classification, resulting in a human-centered framework that integrates data, ML, explainability, visualization, and expert reasoning into an iterative workflow. Finally, the framework was translated into a set of design recommendations that provide practical guidance for the trustworthy development and evaluation of network traffic classification models.

\bmhead{Acknowledgements}

This research work was supported by the National Research Center for Applied Cybersecurity ATHENE. ATHENE is funded jointly by the German Federal Ministry of Research, Technology and Space and the Hessian Ministry of Science and Research, Arts and Culture.

\bmhead{Author Contributions}

Igor Cherepanov was primarily responsible for the conception of the work and the preparation of the manuscript. 
All authors have read and approved the final manuscript and agree to be accountable for all aspects of the work.

\bmhead{Data Availability}

Not applicable.

\bmhead{Code Availability}

Not applicable.

\section{Declarations}
\bmhead{Competing Interests}

The authors have no relevant financial or non-financial competing interests to disclose.

\bmhead{Research Involving Human Participants and/or Animals}

No animals were involved in this research.

\bmhead{Informed Consent}

Participants taking part in this research have done so freely and is based on their willingness to share information crucial for pursuing the purpose and objective of the paper.

\bmhead{Open Access}

This article is licensed under a Creative Commons Attribution 4.0 International License, which permits use, sharing, adaptation, distribution, and reproduction in any medium or format, as long as appropriate credit is given to the original author(s) and the source, a link to the Creative Commons licence is provided, and any changes made are indicated. The images or other third-party material in this article are included in the article's Creative Commons licence, unless indicated otherwise in a credit line to the material. If material is not included in the article's Creative Commons licence and the intended use is not permitted by statutory regulation or exceeds the permitted use, permission must be obtained directly from the copyright holder. To view a copy of this licence, visit \url{https://creativecommons.org/licenses/by/4.0/}.





\bibliography{sn-bibliography}
\end{document}